\documentclass[preprint,superscriptaddress,a4paper,showpacs,showkeys,11pt,nofootinbib]{revtex4}
\usepackage{graphicx} 
\usepackage{graphicx}
\usepackage{amsmath}
\usepackage{booktabs}
\usepackage{siunitx}
\usepackage{url}
\usepackage{hyperref}
\newcommand{\rr}{\mbox{\boldmath $r$}}

\def\pom{{I\!\!P}}

\newcommand{\rb}{\mbox{\boldmath $b$}}

\begin{document}
\title{Photoproduction of $J/\Psi$ in peripheral Oxygen - Oxygen collisions}

\author{Pedro E. A.  da {\sc Costa}}
\email{pazevedo138@gmail.com}
\affiliation{Departamento de F\'isica, Universidade do Estado de Santa Catarina, 89219-710 Joinville, SC, Brazil.}

\author{Andr\'e V. {\sc Giannini}}
\email{AndreGiannini@ufgd.edu.br}
\affiliation{
Federal University of Grande Dourados, 
Faculty of Exact Sciences and Technology,
Zip code 364, 79804-970, Dourados, MS, Brazil}
\affiliation{Departamento de F\'isica, Universidade do Estado de Santa Catarina, 89219-710 Joinville, SC, Brazil.}

\author{Victor P. {\sc Gon\c{c}alves}}
\email{barros@ufpel.edu.br}
\affiliation{Institute of Physics and Mathematics, Federal University of Pelotas (UFPel), \\
  Postal Code 354,  96010-900, Pelotas, RS, Brazil}

\author{Bruno D. {\sc Moreira}}
\email{bduartesm@gmail.com}
\affiliation{Departamento de F\'isica, Universidade do Estado de Santa Catarina, 89219-710 Joinville, SC, Brazil.}

\begin{abstract}
We investigate $J/\Psi$ photoproduction in peripheral oxygen--oxygen ($OO$) collisions at the Large Hadron Collider (LHC), considering different models for the nuclear photon flux, the overlap function, and the dipole--proton scattering amplitude. We present predictions for the corresponding rapidity distributions and total cross sections. Our results indicate that an experimental study of $J/\Psi$ photoproduction in peripheral $OO$ collisions is, in principle, feasible. In addition, they suggest that combining results for this final state in $OO$ collisions with those obtained in $PbPb$ collisions can provide important constraints on the description of photon-induced processes in peripheral collisions.

\end{abstract}

\keywords{$J/\Psi$ photoproduction; Peripheral collisions; QCD dynamics}

\maketitle

\section{Introduction}
One of the main goals of ultrarelativistic heavy - ion collisions is 
the formation of a quark--gluon plasma (QGP), a hot and deconfined state of strongly interacting matter~\cite{Yagi:2005yb}. Over the last decades, the experimental studies performed in $AuAu$ collisions at RHIC and $PbPb$ collisions at the LHC have demonstrated that a  QGP is created in both central ($b \approx 0$) and peripheral ($b < 2R_A$) collisions~\cite{Busza:2018rrf}, where $b$ is the impact parameter of the collision and $R_A$ is the nuclear radius. Understanding  QGP properties and describing the underlying dynamics remain central challenges in Particle Physics. Additionally, the study of ultraperipheral collisions ($b > 2R_A$) \cite{upc}  improved our understanding of photon - induced interactions and enabled the investigation of  distinct final states,that can be used to constrain the description of the QCD dynamics at high energies \cite{klein,gluon,Frankfurt:2001db},  and improve our understanding of the quantum 3D imaging of the partons  inside the protons and nuclei \cite{upc2}. In the last years, several groups have proposed studies  of  QGP properties ~\cite{Nagle:2018nvi,Brewer:2021kiv,Nijs:2021clz} and  photon - induced interactions in  collisions of lighter nuclei~\cite{Goncalves:2022ret, Eskola:2022vaf, Cepila:2025exl}, which are now feasible with the recent oxygen - oxygen ($OO$) collisions at the LHC. The first data have already been released~\cite{CMS:2025bta,CMS:2026qef} and more analyses are expected in the coming years.

\begin{figure}[t]
\includegraphics[scale=0.5]{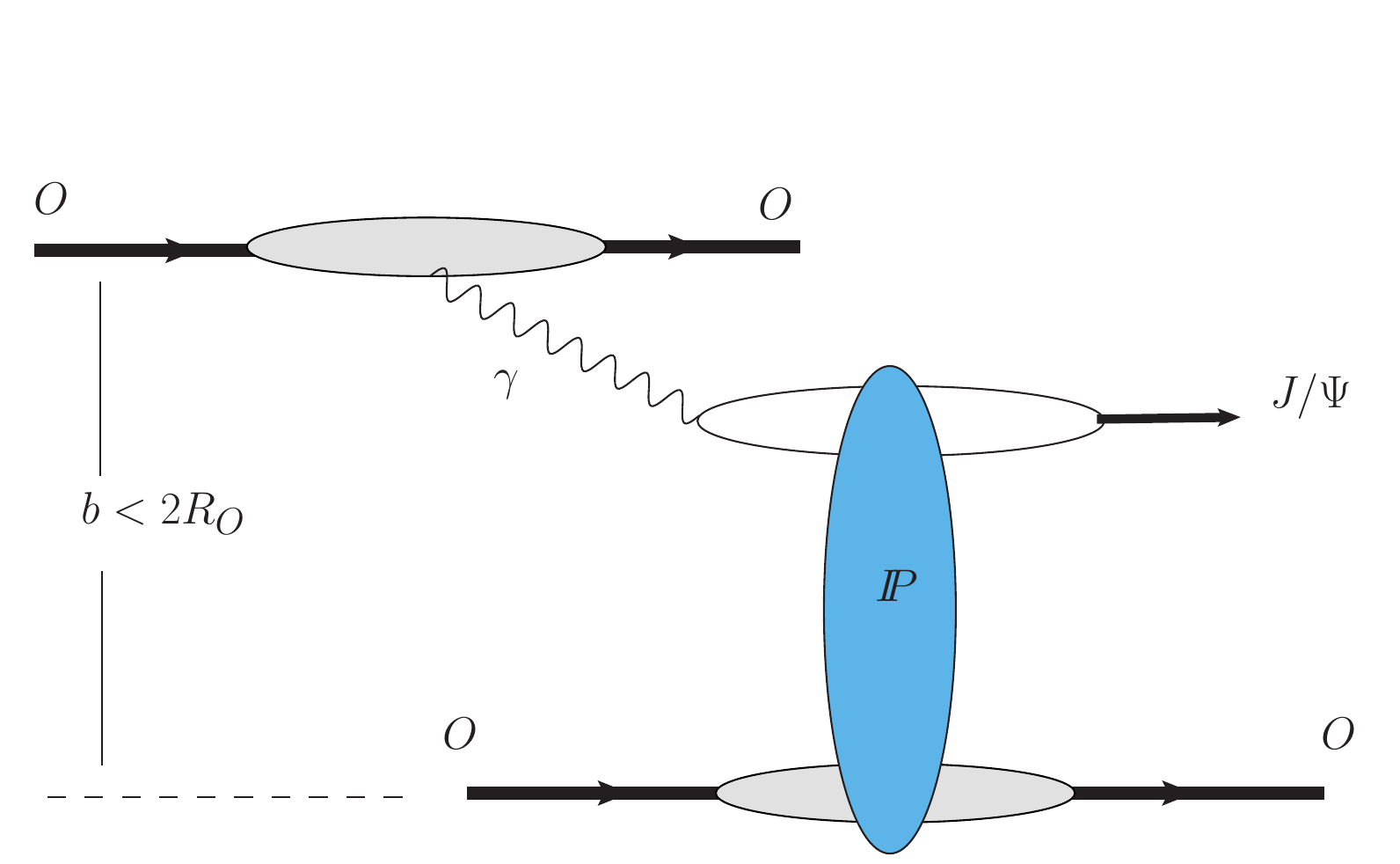}
\caption{Photoproduction of $J/\Psi$ in peripheral $OO$ collisions.}
\label{Fig:diagram}
\end{figure}

The contribution of photon - induced processes in central and peripheral collisions was long expected to be negligible, due to the dominance of strong interactions \cite{upc}. However, in recent years, the STAR,  ALICE and LHCb Collaborations \cite{ALICE:2015mzu,STAR:2019yox,LHCb:2021hoq,ALICE:2022zso,Massacrier:2024fgx,Bize:2024ros} have observed a significant excess in the  yields  $J/\Psi$ mesons and dielectrons at very low transverse momentum in peripheral collisions,  exhibiting features consistent with coherent photon - induced interactions. These observations have motivated the proposition of distinct approaches to extend the framework of photon - induced interactions to peripheral collisions \cite{Klusek-Gawenda:2015hja,Zha:2017jch,GayDucati:2017ksh,Shi:2017qep,Zha:2018ytv,GayDucati:2018who,daCosta:2025frd}. 
At present, the description of 
vector meson photoproduction in peripheral collisions remains an open question. One way to improve our understanding, advocated in Ref.~\cite{daCosta:2025frd}, is a  global analysis of the forthcoming data on $J/\Psi$ and $\Upsilon$ photoproduction in peripheral $PbPb$ collisions. In this paper, we explore a complementary possibility studying   $J/\Psi$ photoproduction in peripheral $OO$ collisions. Since peripheral $PbPb$ collisions were analyzed in detail in Ref.~\cite{daCosta:2025frd}, our goal here is to present, for the first time,  predictions for peripheral $OO$ collisions, using the same ingredients and assumptions as in our previous study.  
 { As we will demonstrate, analyzing the vector meson photoproduction in peripheral $OO$ collisions, represented in Fig.~\ref{Fig:diagram}, allows us to constrain QCD dynamics and the photon flux in a kinematical range complementary to that probed in $PbPb$ collisions.}

{ This paper is organized as follows. In the next section, we briefly review  the formalism used to describe  vector meson photoproduction in peripheral collisions. In particular,  we discuss different approaches for modeling the    oxygen photon flux and analyze  the corresponding impact parameter dependence. For comparison, we also show  results for the  photon flux of a lead nucleus. We then present the model used for the  dipole - nucleus scattering amplitude  and compare predictions for oxygen and lead targets. In section~\ref{sec:res},  the energy dependence of the photoproduction cross-section for   different  impact parameters. We also present our results for the rapidity distributions and total cross-sections for $J/\Psi$  photoproduction in peripheral $OO$ collisions in different centrality classes, obtained with different models for the effective nuclear photon flux, the  dipole - proton scattering amplitude, and  overlap functions.
Results for the ratio of the  $J/\Psi$ photoproduction cross-sections in peripheral $PbPb$ and $OO$ collisions are also presented. Finally, in section~\ref{sec:sum}, we summarize our main results and conclusions.

}


\section{Formalism}

In what follows, we will present a brief review of the formalism used to describe the   $J/\Psi$ photoproduction  in peripheral $OO$ collisions at the LHC, represented in Fig. \ref{Fig:diagram}, highlighting the main ingredients of our calculation.   Assuming the validity of the  equivalent photon approximation (EPA) \cite{epa}, the differential cross - section  can be written as
\begin{eqnarray}
\frac{d\sigma \,\left[O O \rightarrow   O \otimes  J/\Psi \otimes O\right]}{d^2\rb \, dy_{\Psi}} = \omega_1 N(\omega_1,b)\,\sigma_{\gamma O \rightarrow J/\Psi \otimes O}\left(\omega_1 \right) + (\omega_1 \rightarrow  \omega_2) \,\,,
\label{dsigdy}
\end{eqnarray}
where  $\rb$ is the impact parameter ($b = |\rb|)$, $y_{\Psi}$ is the rapidity  of the produced vector meson, and $\sigma_{\gamma O \rightarrow V \otimes O}$ is the cross - section for the  $J/\Psi$ production in a photon - oxygen interaction. The symbol $\otimes$ indicates that the photon - nucleus interaction is mediated by a color singlet object, usually denoted as Pomeron $\pom$, which implies that a rapidity gap is expected to be present in the final state. Moreover, $\omega_1 = (m_{\Psi}/2)e^{y_{\Psi}}$ and $\omega_2 = (m_{\Psi}/2)e^{-y_{\Psi}}$, with $m_{\Psi}$ the vector meson mass. Finally, $N(\omega,b)$ is the photon spectrum, which can be expressed  as  \cite{upc}
\begin{eqnarray}
 N(\omega,b) = \frac{Z^{2}\alpha}{\pi^2}\frac{1}{b^{2} v^{2}\omega}
\cdot \left[
\int u^{2} J_{1}(u) 
F\left(
 \sqrt{\frac{\left( \frac{b\omega}{\gamma_L}\right)^{2} + u^{2}}{b^{2}}}
 \right )
\frac{1}{\left(\frac{b\omega}{\gamma_L}\right)^{2} + u^{2}} {d}u
\right]^{2} \,\,,
\label{Eq:fluxo0}
\end{eqnarray}
where $\alpha$ is the electromagnetic coupling constant, $\gamma_L$ is the Lorentz factor,  $v$ is the nuclear velocity, and   $F(q)$ is the charge form factor. 
{ For a spherical nucleus, the relation between the form factor and the charge density $\rho(r)$ is given by  
\begin{eqnarray}
    F(q^2) &= \frac{4\pi}{qA}\int r\,\rho(r)\,\sin(qr)\,dr \,\,.
\end{eqnarray}
In  our analysis, we compute the photon flux using  the three-parameter Woods - Saxon charge density, 
\begin{eqnarray}
\rho(r) = \dfrac{\rho_0\left(1+\dfrac{\kappa r^2}{R_A^2}\right)}{1+\exp\left(\dfrac{r-R_A}{a}\right)} \,\,,
\end{eqnarray}
where $\rho_0$ is the central density, $a$ is the skin depth, and $\kappa$ is the deformation parameter, which accounts for deviations from a smooth spherical surface.
For oxygen, we take \(\kappa = -0.051, R_A = 2.608\,\text{fm}\), and  \(a = 0.513\,\text{fm}\)~\cite{DeJager:1974liz}. The description of the Oxygen density profile is still under debate (see, e.g., Ref.~\cite{Loizides:2025ule}), with the main differences between  models occurring at small $r$. We have verified that, for the centralities considered in this study (50\% - 70\% and 70\% - 90\%), the impact of a alternative choices for $\rho(r)$ is small, but it becomes non-negligible for more central collisions. We leave a detailed discussion of this issue for future work.     \begin{figure}[t]
 	\centering
 	\includegraphics[width=0.45\linewidth]{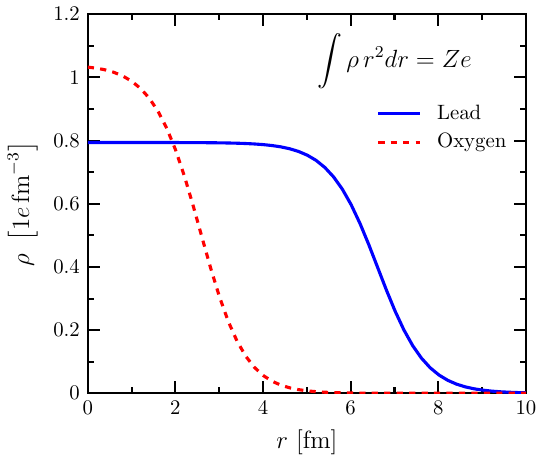}
    \includegraphics[width=0.47\linewidth]{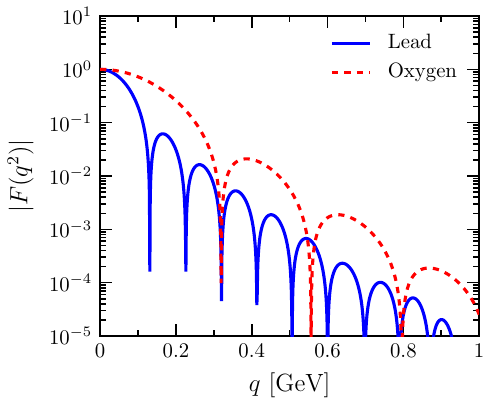}
        \caption{Woods-Saxon density distributions (left panel) and associated nuclear form factors (right panel) for the oxygen and lead nuclei.}
        \label{fig:form_factor}
    \end{figure}
  In Fig.~\ref{fig:form_factor} (left panel) we show  the radial dependence of the Woods-Saxon density distribution. For comparison, we also show the distribution for a lead nucleus, obtained with \( \kappa = 0, R_A = 6.62\,\text{fm}\) and  \(a = 0.549\,\text{fm}\).
These different radial distributions for oxygen and lead imply that the   corresponding form factors (right panel)  have different momentum dependence. As we show below, these differences  have direct implications for the impact parameter dependence of the associated photon fluxes.

}

Our focus in this paper is on peripheral $OO$ collisions, where the impact parameter between the incident ions is smaller than the sum of their nuclear radii ($b < 2 R_O$). A transverse - plane view of  $J/\Psi$ photoproduction in these collisions is shown in Fig. \ref{colisao_geometria}. The geometry collision is characterized by three vectors: (a) the ion - ion impact parameter $|\rb|$; (b) the distance from the center of the ion that has emitted the photon and the point at which this photon interacts with the ion target: $|\rb_1|$; and (c) the distance from the center of the target ion to the interaction point: $|\rb_2| = |\rb_1 - \rb|$. The modulus of these vectors are related by ${b}_2^2 = {b}_1^2 + {b}^2 - 2b_1 b \cos(\beta)$.  Following Ref.~\cite{daCosta:2025frd}, we  consider four different models to extend the treatment of the nuclear photon flux to peripheral heavy - ion collisions. Denoting by $A$ the nucleus that is the photon emitter and by $B$ the target nucleus, the different models for the effective photon flux can be expressed by:
\begin{enumerate}
\item 
\begin{equation}
N_A^{(0)}(\omega, b) = N_A(\omega, b) \,\,\,,
\end{equation}
 i.e.  the nuclear photon flux is not modified in peripheral collisions;
\item \begin{equation}
N^{(1)}_A(\omega, b) = \int N^{(0)}_A(\omega, {b}_1) \frac{\theta(R_B - b_2)}{\pi R_B^2} \,d^2b_1\,\,,
\label{Eq:flux1} 
\end{equation}
     \begin{figure}[t]
 	\centering
 	\includegraphics[width=0.45\linewidth]{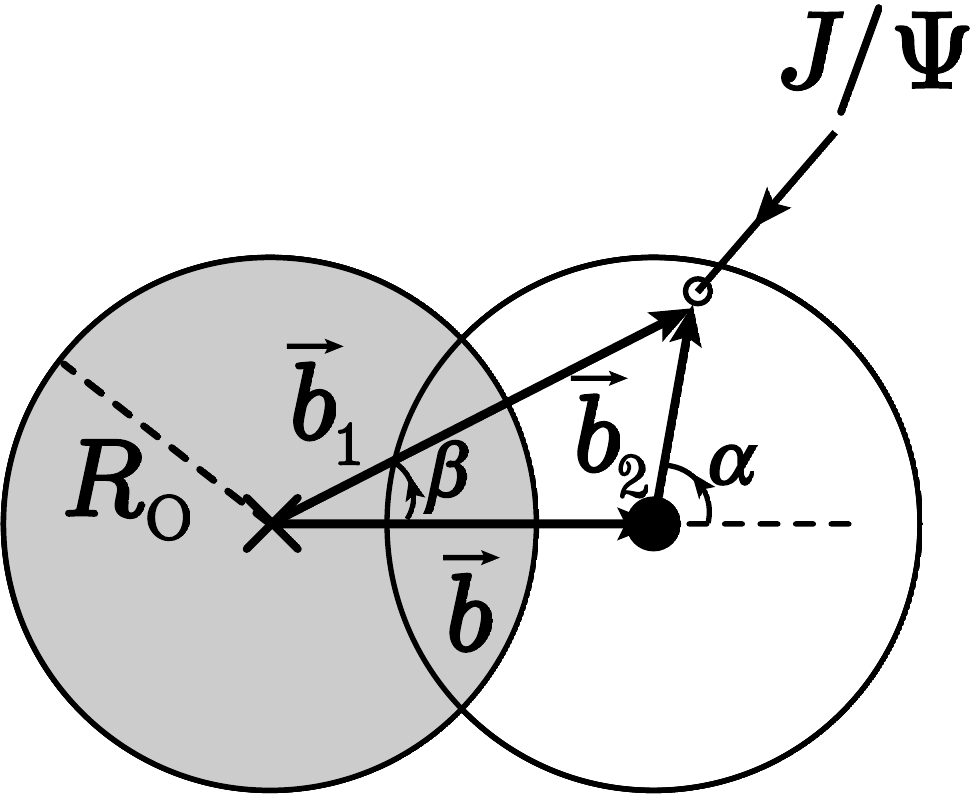}
        \caption{Transverse - plane view of the $J/\Psi$ photoproduction in a peripheral $OO$ collision.}
        \label{colisao_geometria}
    \end{figure}
where the theta function ensures that the photon emitted by  one of the nucleus hits the second nucleus. Such a model considers that the vector meson photoproduction only occurs inside the nuclear target and the overlapping region between the incident ions is included as a possible interaction region \cite{Klusek-Gawenda:2015hja};
\item \begin{equation}
	N^{(2)}_A(\omega, b) = \int N_A^{(0)}(\omega, {b}_1) \frac{\theta(R_B - b_2) \times \theta(b_1 - R_A)}{\pi R_B^2} d^2b_1 \,\,.
    \label{Eq:flux2} 
    \end{equation}
    In this model, the overlapping region is excluded in the modeling of the effective nuclear photon flux, which is motivated by the fact that the photoproduction of vector mesons in this region can be strongly affected by the presence of a nuclear or hot medium \cite{Klusek-Gawenda:2015hja};
\item  \begin{equation}
	N^{(3)}_A(\omega, b) = \int N_A^{(0)}(\omega, b_1) \frac{\theta(R_B - b_2) \times \theta(b_1 - R_A)}{A_{\text{eff}}(b)} \,d^2b_1 \,\,,
    \label{Eq:flux3} 
    \end{equation}  
with   
\begin{equation}
	A_{\text{eff}}(b) = R_B^2 \left[\pi - 2\arccos\left(\frac{b}{2R_B}\right)\right] + \frac{b}{2} \sqrt{4R_B^2 - b^2} \,\,.
\end{equation} 
In this model the constant factor $\pi R_B^2$ in the denominator of Eq.  (\ref{Eq:flux2}) is generalized by an effective area, which depends on the impact parameter of the collision \cite{GayDucati:2018who}. 
\end{enumerate}

     \begin{figure}[t]
 	\centering
 	\includegraphics[width=0.45\linewidth]{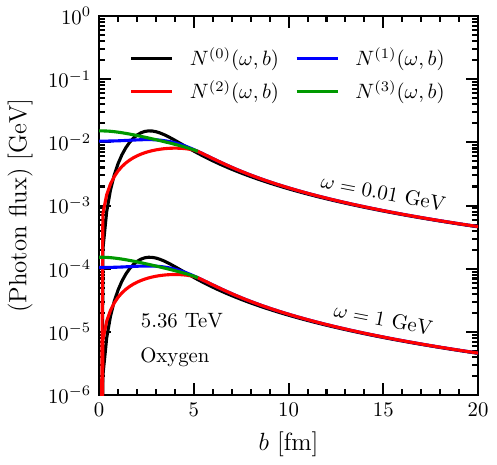}
 	\includegraphics[width=0.45\linewidth]{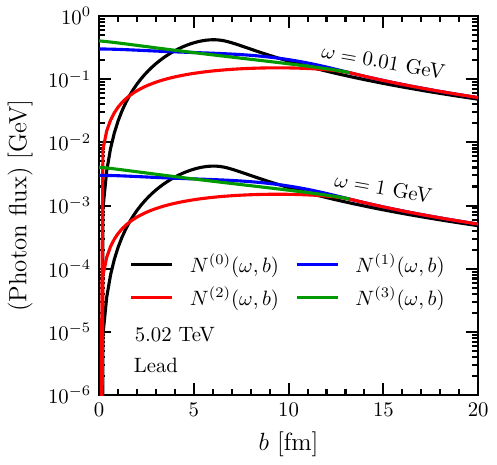}
        \caption{Impact parameter dependence of the distinct models for the photon flux associated with an Oxygen (left panel) and a Lead (right panel) nucleus. Results for distinct values of the photon energy.}
        \label{fig:photon_flux}
    \end{figure}

{ 
In Fig.~\ref{fig:photon_flux} (left panel) we present  the impact parameter dependence of the distinct models for the Oxygen photon flux, derived considering two values for the photon energy $\omega$. For comparison, we present in the right panel the results for the photon flux associated with a Lead nucleus. As expected, the  distinct models predict the same behavior for large values of $b$ and have a larger normalization for a heavier nuclei. On the other hand, the behavior for small-$b$ is strongly model dependent, with the range of impact parameters where the models differ being dependent of the atomic number.  As a consequence,  the dependence on the modeling of the nuclear flux in  peripheral $OO$ collisions is expected to be different from that for  $PbPb$ one.
}

\begin{figure}[t]
 	\centering
 	\includegraphics[width=0.85\linewidth]{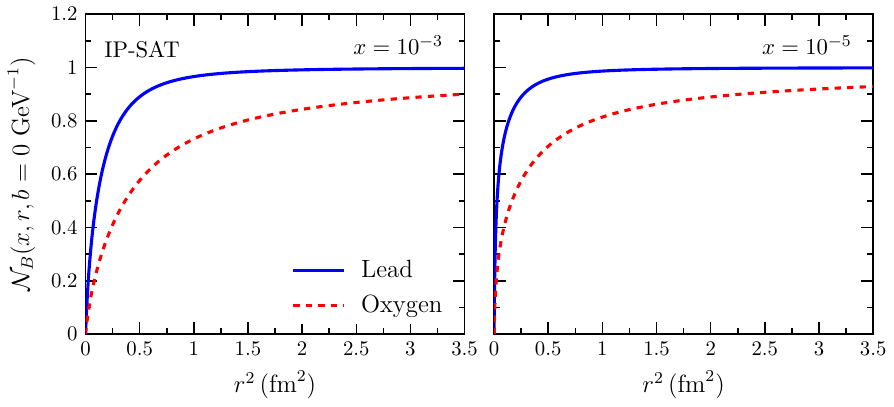}
        \caption{Dipole-nucleus scattering amplitude as a function of the dipole size squared $r^2$ for two distinct values of $x$ and central collisions ($\rb =0$).}
        \label{fig:ene_nuclear}
    \end{figure}

In our analysis we will also assume that only the nucleons of the nuclear target that are not in the overlapping region  contribute for the interaction, i.e. that only spectator nucleons act as target, which implies that  the  nuclear vector  photoproduction cross - section   will be expressed in the color dipole formalism by~\cite{GayDucati:2018who}
\begin{equation}
\sigma_{\gamma B \rightarrow V B}(W,b) = \int d{^2\rb_2}\, \theta(b_1 - R_A) \left\{ \int \int dz \, d^2\rr  [\psi^*_{\Psi}(r,z)\psi(r,z)]_T\, \mathcal{N}_B(x, \rr, \rb_2) \right\}^2\,\,,
\label{Eq:sec_bdep}
\end{equation} 
where $W = [m_{\Psi}\sqrt{s_{NN}}\exp(-y_{\Psi})]^{1/2}$ is photon - nucleus center - of - mass energy,   $z$ $(1-z)$ is the
longitudinal momentum fractions of the quark (antiquark),  $(\psi_{\Psi}^{*}\psi)_{T}$ denotes the overlap of the transverse photon and $J/\Psi$ wave functions, and $x = W^2/m_{\Psi}^2$. As in Ref.~\cite{daCosta:2025frd}, we will present the predictions for two distinct models for the overlap function, the Boosted Gaussian (BG) and the Gaus-LC (GLC) models, which differ in the treatment of the scalar part of the vector meson wave function (For a more detailed discussion, see section II.B in Ref.~\cite{daCosta:2025frd}). Moreover, 
 ${\cal{N}}_{B} (x,\rr,\rb)$ denotes the non-forward scattering  amplitude of a dipole of size $\rr$ on the target $B$, which is  directly related to  the QCD dynamics \cite{hdqcd}.
  {We will estimate ${\cal{N}}_B$  assuming the Glauber-Gribov (GG) formalism~\cite{glauber,gribov,mueller,Armesto:2002ny}, which allow us to express this quantity in terms of the  dipole-proton cross-section ($\sigma_{dp}$), as follows
  \begin{eqnarray}
{\cal{N}}_B(x,\rr,\rb) =  1 - \exp \left[-\frac{1}{2}  \, \sigma_{dp}(x,\rr^2) \,T_B(\rb)\right] \,,
\label{enenuc}
\end{eqnarray}
where the nuclear profile function $T_B(\rb)$ is described by a Wood-Saxon distribution. The dipole-proton cross section, $\sigma_{dp}$, is expressed in terms of the dipole-proton scattering amplitude as follows
\begin{eqnarray}
\sigma_{dp}(x,\rr^2) = 2 \int d^2\rb_p \, {\cal{N}}_p(x,\rr,\rb_p) \,,
\end{eqnarray}
with  $\rb_p$ being the impact - parameter for the dipole-proton interaction. }  
  In what follows, we will perform our calculations considering the bCGC~\cite{KMW,Watt_bCGC}, IP-SAT~\cite{ipsat1,ipsat3,ipsat4,ipsat_heikke} and IPnon-SAT~\cite{ipsat_heikke} models for the description of ${\cal{N}}_p$. These models differ in the treatment of nonlinear effects on the QCD dynamics at the proton level. While the bCGC and IP-SAT take into account of these effects, considering distinct approaches, the IPnon-SAT model disregards the nonlinear effects (For details, see Ref.~\cite{daCosta:2025frd}).  
{ In Fig.~\ref{fig:ene_nuclear}, we present the nuclear scattering amplitude ${\cal{N}}_B(x,\rr,\rb)$ as a function of $r^2$ for a Oxygen target considering two values of the Bjorken - $x$ variable and $b = 0$, derived assuming the IP-SAT model as input of the calculations.  For comparison, we also present the results for the dipole - Lead scattering amplitude. One has that for both nuclei ${\cal{N}}_B \le 1$, satisfying unitarity. However, the $r^2$-dependence is
dependent on the atomic number,  with the onset of the saturation regime (${\cal{N}}_B \approx 1$) being slower in the Oxygen case.  In other words, the transition between the linear
and non - linear regimes of the QCD dynamics is dependent on $A$, with  oxygen targets providing the opportunity to probe the saturation effects in a new kinematical regime.
}

     \begin{figure}[t]
 	\centering
 	\includegraphics[width=0.45\linewidth]{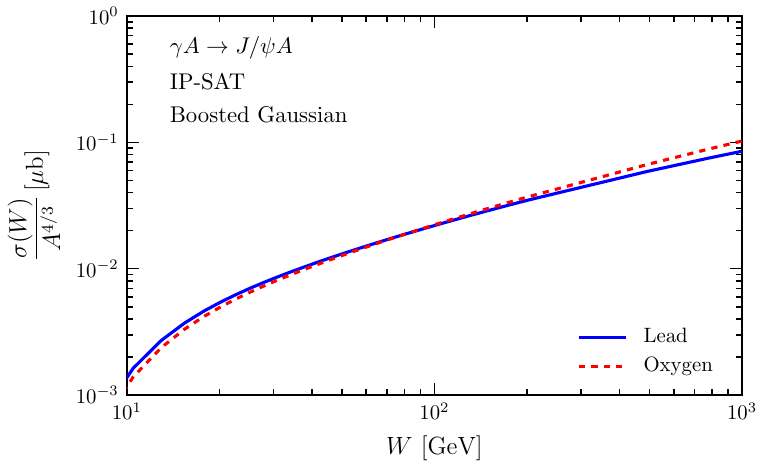}
 	\includegraphics[width=0.45\linewidth]{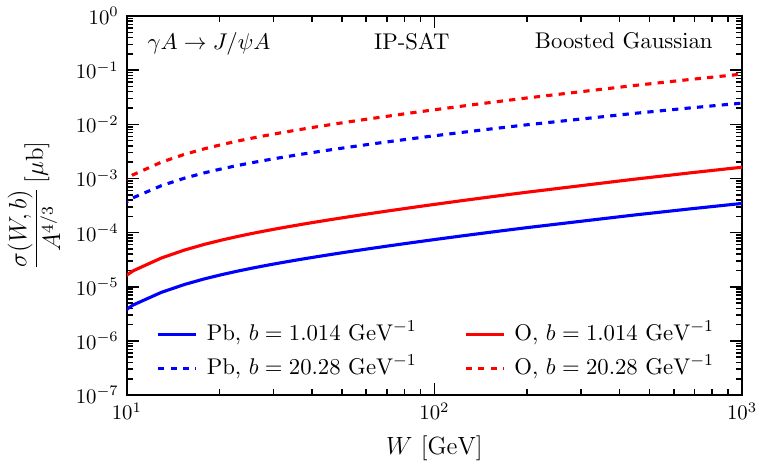}
        \caption{Energy dependence of the cross-section, scaled by a factor $A^{4/3}$, for the $J/\Psi$ production in photon - nucleus interactions. Results for the Oxygen and Lead targets. {\it Left panel:} Predictions for the $b$ - independent cross-section, $\sigma_{\gamma A \rightarrow V A}(W)$, derived disregarding the theta function in Eq. (\ref{Eq:sec_bdep}). {\it Right panel:} Predictions for the $b$ - dependent cross-section, $\sigma_{\gamma A \rightarrow V A}(W,b)$, for different values of the impact parameter and distinct nuclei.}
        \label{fig:crossVM}
    \end{figure}

\begin{figure}[t]
	\centering
	\includegraphics[width=0.8\textwidth]{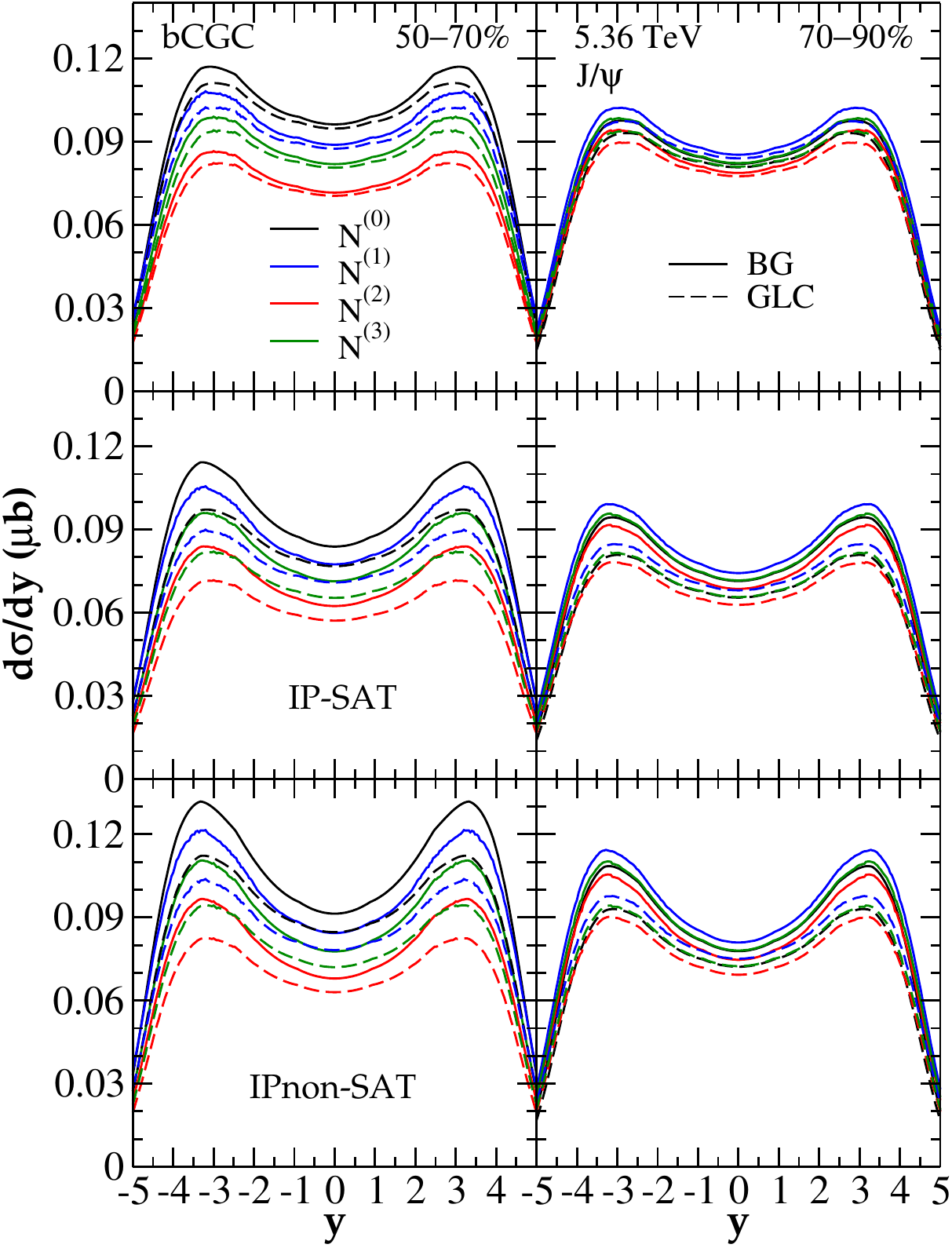}
	\caption{Predictions for the rapidity distributions associated with the photoproduction of $J/\Psi$ mesons in peripheral $OO$ collisions at the centralities 50 - 70\% and 70 - 90\%, derived assuming distinct models for the effective nuclear photon flux,  dipole - proton scattering amplitude and  overlap functions. Results for  \( \sqrt{s_{NN}} = 5.36\,\mathrm{TeV} \).  }
    \label{Fig:jpsiDSIGDY_HC}
\end{figure}


\section{Results}
\label{sec:res}

{ Initially, let's analyze the energy dependence of the photon - nucleus cross-section, $\sigma_{\gamma A \rightarrow V A}$, for different values of the atomic number. We will scale our predictions by a factor $A^{4/3}$, which is the nuclear dependence expected in the absence of saturation effects \cite{Mantysaari:2017slo}. The corresponding results for  oxygen and lead nuclei
are presented in Fig.~\ref{fig:crossVM} (left panel). These results were obtained by neglecting  the theta function in Eq. (\ref{Eq:sec_bdep}) and using the IP-SAT and Boosted Gaussian models as inputs.   We find similar behaviors at low energies. At high energies, however, we predict a steeper rise for an oxygen target. This behavior is expected, since  non-linear effects are stronger for heavier nuclei. It is important to emphasize that in our approach to  peripheral nuclear collisions the relevant quantity is the impact-parameter-dependent cross - section $\sigma_{\gamma B \rightarrow V B}(W,b)$, given by Eq.(\ref{Eq:sec_bdep}), which  is sensitive to the nuclear geometry and, therefore, exhibits a stronger dependence on the atomic number. This expectation is confirmed in the right panel of Fig.~\ref{fig:crossVM}, where we compare the  energy dependence of the cross-sections at two fixed values of the impact parameter. This difference has a direct impact on the predictions for  $J/\Psi$ photoproduction in peripheral $OO$ collisions compared with those for  $PbPb$.}

In what follows, we  present our predictions for the rapidity distributions associated with  $J/\Psi$ photoproduction  in peripheral $OO$ collisions  at the LHC energy of \( \sqrt{s_{NN}} = 5.36\,\mathrm{TeV} \). { As in Ref.~\cite{daCosta:2025frd}, we assume $m_c = 1.27/\,1.3528/\,1.3504$ GeV in  calculations using the bCGG / IP-SAT / IPnon - SAT models, respectively}. The results are shown for different centralities and for distinct combinations of  models for the nuclear photon flux,  overlap function, and dipole - proton scattering amplitude, in order  to estimate the current theoretical uncertainty in the description of peripheral $OO$ collisions. The predictions for the different centralities  are obtained assuming  that the centrality $c$ and the impact parameter  $b$ are related by  $c = {b^2}/({4R_A^2})$, where $R_A$ is the nuclear radius. { Our main motivation for using this simplified geometric definition of  centrality is to facilitate comparisons with previous studies~\cite{Klusek-Gawenda:2015hja, GayDucati:2018who,daCosta:2025frd}, which adopted the  same definition.  However, it is important to emphasize that this relation provides only  a rough estimate of the collision centrality; more realistic predictions can be obtained within a Glauber approach \cite{Loizides:2017ack,Loizides:2025ule}.
Such distinct approaches to determine the centrality imply distinct
values for the minimum and maximum impact parameters associated with a given centrality range, with the discrepancies increasing for more central collisions. To minimize the sensitivity of our predictions to the centrality definition, and to the modeling of the nuclear form factor, we  focus on the  following peripheral centrality classes:  50\% - 70\% and 70\% - 90\%.} { Another motivation to restrict our analysis to these classes is related to the fact,  already emphasized in Ref.~\cite{daCosta:2025frd}, that in our view the impact of  QGP formation on the photoproduced  vector mesons cannot be neglected in more central collisions.
An interesting aspect is that the impact of a QGP on the  $J/\Psi$ photoproduction of is expected to  differ in $PbPb$ and $OO$ collisions, since the  properties of the QGP produced in these systems are predicted to be different (See, e.g, Refs. \cite{Nagle:2018nvi,Brewer:2021kiv,Nijs:2021clz}). Consequently, the study of  $J/\Psi$ photoproduction in more central nuclear collisions   could, in principle,  be used to probe the QGP. A detailed analysis of photon - induced processes in more central collisions will be presented in a forthcoming publication.
}

\begin{table}[t]
\centering
 \centering
 \renewcommand{\arraystretch}{1.6}
 \begin{tabular}{||l|c|c|c|c||}
 \hline\hline
\multicolumn{3}{c}{$\sigma^{J/\psi}_{\text{tot}} (\mu b) \,\,\,\,\,\,\,(-2 \leq y_{\Psi} \leq 2)$} \\
 \hline
 &\multicolumn{2}{c||}{\textbf{5.36 TeV}} \\
 \hline
\textbf{Dipole - proton model}  & \textbf{50\%--70\%} & \textbf{70\%--90\%} \\
 \hline\hline
\multicolumn{3}{c}{Effective nuclear photon flux: $N^{(0)}(\omega,b)$} \\
 \hline
bCGC &            0.400 (0.391) &      0.340 (0.333)  \\
IP-SAT &           0.355 (0.321) &      0.303 (0.274)  \\
IPnon-SAT &        0.391 (0.357) &      0.333 (0.304)  \\
 \hline\hline
\multicolumn{3}{c}{Effective nuclear photon flux: $N^{(1)}(\omega,b)$} \\
 \hline
bCGC &         0.368 (0.360) &      0.354 (0.346)  \\
IP-SAT &       0.327 (0.296) &      0.315 (0.285)  \\
IPnon-SAT &    0.360 (0.329) &      0.346 (0.317)  \\
 \hline\hline
\multicolumn{3}{c}{Effective nuclear photon flux: $N^{(2)}(\omega,b)$} \\
 \hline
bCGC &         0.298 (0.291) &      0.327 (0.320)  \\
IP-SAT &       0.265 (0.239) &      0.291 (0.263)  \\
IPnon-SAT &     0.291 (0.266) &      0.320 (0.292)  \\
 \hline\hline
\multicolumn{3}{c}{Effective nuclear photon flux: $N^{(3)}(\omega,b)$} \\
 \hline
bCGC &         0.340 (0.333) &      0.342 (0.334)  \\
IP-SAT &       0.303 (0.274) &      0.304 (0.275)  \\
IPnon-SAT &    0.333 (0.304) &      0.334 (0.305)  \\
 \hline\hline
 \end{tabular}
\caption{Predictions for the total cross-sections associated with the photoproduction of $J/\Psi$ mesons in  peripheral $OO$ collisions at central rapidities ($-2 \leq y_{\Psi} \leq 2$) and \( \sqrt{s} = 5.36\,\mathrm{TeV} \). Results for different centralities,  derived considering distinct models for the effective nuclear photon flux,  dipole - proton scattering amplitude and overlap function. The predictions in parentheses are for the GLC model.}
\label{Tab:central}
\end{table}

\begin{table}[t]
 \centering
 \renewcommand{\arraystretch}{1.6}
 \begin{tabular}{||l|c|c|c|c||}
 \hline\hline
\multicolumn{3}{c}{$\sigma^{J/\psi}_{\text{tot}} (\mu b) \,\,\,\,\,\,\,(2 \leq y_{\Psi} \leq 4.5)$} \\
 \hline
 &\multicolumn{2}{c||}{\textbf{5.36 TeV}} \\
 \hline
\textbf{Dipole - proton model} &  \textbf{50\%--70\%} & \textbf{70\%--90\%} \\
 \hline\hline
\multicolumn{3}{c}{Effective nuclear photon flux: $N^{(0)}(\omega,b)$} \\
 \hline
bCGC &         0.265 (0.251) &      0.216 (0.205)  \\
IP-SAT &       0.259 (0.221) &      0.210 (0.180)  \\
IPnon-SAT &    0.298 (0.255) &      0.242 (0.207)  \\
 \hline\hline
\multicolumn{3}{c}{Effective nuclear photon flux: $N^{(1)}(\omega,b)$} \\
 \hline
bCGC &         0.245 (0.232) &      0.229 (0.218)  \\
IP-SAT &       0.239 (0.204) &      0.224 (0.191)  \\
IPnon-SAT &    0.275 (0.235) &      0.257 (0.220)  \\
 \hline\hline
\multicolumn{3}{c}{Effective nuclear photon flux: $N^{(2)}(\omega,b)$} \\
 \hline
bCGC &         0.194 (0.184) &      0.211 (0.200)  \\
IP-SAT &       0.190 (0.162) &      0.205 (0.175)  \\
IPnon-SAT &    0.218 (0.187) &      0.236 (0.203)  \\
 \hline\hline
\multicolumn{3}{c}{Effective nuclear photon flux: $N^{(3)}(\omega,b)$} \\
 \hline
bCGC &         0.222 (0.211) &      0.220 (0.209)  \\
IP-SAT &       0.217 (0.185) &      0.215 (0.183)  \\
IPnon-SAT &    0.249 (0.214) &      0.247 (0.212)  \\
 \hline\hline
 \end{tabular}
\caption{Predictions for the total cross-sections associated with the photoproduction of $J/\Psi$ mesons in  peripheral $OO$ collisions at forward rapidities ($2 \leq y_{\Psi} \leq 4.5$) and \( \sqrt{s} = 5.36\,\mathrm{TeV} \). Results for different centralities,  derived considering distinct models for the effective nuclear photon flux,  dipole - proton scattering amplitude and overlap function. The predictions in parentheses are for the GLC model.}
\label{Tab:forward}
\end{table}

In Figs. \ref{Fig:jpsiDSIGDY_HC}   we present our predictions for the rapidity distributions associated with the photoproduction of $J/\Psi$ mesons in peripheral $OO$ collisions at  \( \sqrt{s} = 5.36\,\mathrm{TeV} \) and different centralities,  assuming distinct models for the effective nuclear photon flux,  dipole - proton scattering amplitude and  overlap function. Our results indicate the following  regarding the description of the:
\begin{itemize}
    \item {\it Overlap function:} The  BG model for the overlap function implies a larger  normalization, with the impact being dependent on the centrality;
    \item {\it Effective nuclear photon flux:} For the centrality of 70 \% - 90 \%, the upper (lower) bound is provided by the $N^{(1)}$ ($N^{(2)}$) model. In contrast, for smaller centralities,  the $N^{(0)}$ model for the effective nuclear photon flux provides an upper  bound for the predictions;
    \item {\it QCD dynamics in the dipole - proton cross - section:} The  bCGC model provides an upper bound at central rapidities, with the IP-SAT and IPnon-SAT results being similar. { The IP-SAT and IPnon-SAT predictions become more distinct at forward/backward rapidities, when smaller values of the Bjorken-$x$ variable are probed in comparison to midrapidities.} In addition, the difference between the predictions, derived assuming distinct models for $\sigma_{dp}$, is dependent on the centrality.
\end{itemize}
For completeness, in Tables   \ref{Tab:central}  and \ref{Tab:forward} we present the corresponding predictions for the cross - sections integrated over distinct rapidity ranges. Considering the integrated luminosity delivered by LHC during the $OO$ run, ${\cal{O}}(6 \, \mbox{nb}^{-1})$, these results indicate that an experimental analysis of this process using the collected data is, in principle, feasible. Moreover, they point out that  experimental data for different centralities and rapidity ranges will provide important constraints on the modeling of peripheral $OO$ collisions.

In Ref.~\cite{daCosta:2025frd} we  proposed a  simultaneous analysis of different heavy meson final states, such as $J/\Psi$ and $\Upsilon$, in  $PbPb$ collisions at a fixed center - of - mass energy, as a way to 
disentangle  the different models and assumptions used in modeling of photoproduction processes in peripheral collisions.
The study of $J/\Psi$ photoproduction in peripheral $OO$ collisions offers a complementary way to pursue this goal, by analyzing the same final state in different colliding systems.  Here, we { perform a first exploratory investigation and} estimate the ratio between  $J/\Psi$ photoproduction cross sections in peripheral $PbPb$ and $OO$ collisions, computed using the same ingredients. { In particular, we  assume the simplified geometric definition of centrality in both cases. This is a strong assumption and should be  improved in  future studies.   } For $PbPb$ collisions, we estimate the cross - sections at $\sqrt{s_{NN}} = 5.02 \,\text{TeV}$ using the approach of Ref.~\cite{daCosta:2025frd}. Note that we are therefore taking the ratio of cross-sections computed at different values of $\sqrt{s_{NN}}$, since we intend to derive reflecting the recent $PbPb$ and $OO$ runs. { Our motivation for studying  this ratio is that the dependence on the modeling of the overlap function and the dipole - nucleus scattering amplitude is expected to be reduced,which may help to constrain the effective nuclear photon flux in peripheral collisions.} In Tables \ref{Tab:ratio_central} and \ref{Tab:ratio_forward}, we present our predictions for the ratio
$\sigma^{J/\psi}_{\text{PbPb}}(\sqrt{s_{NN}} = 5.02 \,\text{TeV}) \,/\, \sigma^{J/\psi}_{\text{OO}}(\sqrt{s_{NN}} = 5.36 \,\text{TeV})$ for the rapidity ranges covered by a central detector $(-2 \leq y_{\Psi} \leq 2)$ and a forward detector  $(2 \leq y_{\Psi} \leq 4.5)$, respectively.  Our findings suggest that this ratio is primarily sensitive to the choice of effective nuclear photon flux, rather than to dipole-proton scattering amplitude or the overlap function. Consequently, future experimental measurements could be instrumental in discriminating among competing theoretical approaches to peripheral collisions.

 \begin{table}[t]
 \centering
 \renewcommand{\arraystretch}{1.6}
 \begin{tabular}{||l|c|c|c|c||}
 \hline\hline
\multicolumn{3}{c}{$\sigma^{J/\psi}_{\text{PbPb}}(\sqrt{s_{NN}} = 5.02 \,\text{TeV}) \,/\, \sigma^{J/\psi}_{\text{OO}}(\sqrt{s_{NN}} = 5.36 \,\text{TeV}) \,\,\,\,\,\,\, (-2 \leq y_{\Psi} \leq 2)$} \\
 \hline
\textbf{Dipole - proton model}  & \textbf{50\%--70\%} & \textbf{70\%--90\%} \\
 \hline\hline
\multicolumn{3}{c}{Effective nuclear photon flux: $N^{(0)}(\omega,b)$} \\
 \hline
bCGC &        3348.93 (3181.63) &    3127.22 (2972.31)  \\
IP-SAT &       3393.27 (3219.27) &    3167.35 (3006.88)  \\
IPnon-SAT &      3311.39 (3119.67) &    3091.21 (2914.09)  \\
 \hline\hline
\multicolumn{3}{c}{Effective nuclear photon flux: $N^{(1)}(\omega,b)$} \\
 \hline
bCGC &        3920.68 (3724.41) &    3502.97 (3329.27)  \\
IP-SAT &      3973.47 (3769.03) &    3548.39 (3368.35)  \\
IPnon-SAT &     3877.94 (3652.67) &    3463.38 (3264.48)  \\
 \hline\hline
\multicolumn{3}{c}{Effective nuclear photon flux: $N^{(2)}(\omega,b)$} \\
 \hline
bCGC &        3693.51 (3509.04) &    3387.15 (3219.13)  \\
IP-SAT &      3742.30 (3550.48) &    3430.91 (3256.79)  \\
IPnon-SAT &      3651.94 (3440.62) &    3348.56 (3156.45)  \\
 \hline\hline
\multicolumn{3}{c}{Effective nuclear photon flux: $N^{(3)}(\omega,b)$} \\
 \hline
bCGC &       3695.26 (3510.67) &    3386.82 (3218.82)  \\
IP-SAT &       3744.13 (3552.16) &    3430.61 (3256.49)  \\
IPnon-SAT &     3653.69 (3442.22) &    3348.23 (3156.05)  \\
 \hline\hline
 \end{tabular}
\caption{Predictions for the ratio between total cross-sections associated with the photoproduction of $J/\Psi$ mesons in  peripheral $PbPb$ collisions at \( \sqrt{s} = 5.02\,\mathrm{TeV} \) and $OO$ collisions at  \( \sqrt{s} = 5.36\,\mathrm{TeV} \). Results for central rapidities ($-2 \leq y_{\Psi} \leq 2$) and different centralities,  derived considering distinct models for the effective nuclear photon flux,  dipole - proton scattering amplitude and overlap function. The predictions in parentheses are for the GLC model.}
\label{Tab:ratio_central}
 \end{table}

 \begin{table}[t]
 \centering
 \renewcommand{\arraystretch}{1.6}
 \begin{tabular}{||l|c|c|c|c||}
 \hline\hline
\multicolumn{3}{c}{$\sigma^{J/\psi}_{\text{PbPb}}(\sqrt{s_{NN}} = 5.02 \,\text{TeV}) \,/\, \sigma^{J/\psi}_{\text{OO}}(\sqrt{s_{NN}} = 5.36 \,\text{TeV}) \,\,\,\,\,\,\, (2 \leq y_{\Psi} \leq 4.5)$} \\
 \hline
\textbf{Dipole - proton model}  & \textbf{50\%--70\%} & \textbf{70\%--90\%} \\
 \hline\hline
\multicolumn{3}{c}{Effective nuclear photon flux: $N^{(0)}(\omega,b)$} \\
 \hline
bCGC &        1950.02 (1881.14) &    1739.47 (1683.01)  \\
IP-SAT &       1950.49 (1902.63) &    1734.13 (1698.42)  \\
IPnon-SAT &     1861.53 (1811.29) &    1654.18 (1616.14)  \\
 \hline\hline
\multicolumn{3}{c}{Effective nuclear photon flux: $N^{(1)}(\omega,b)$} \\
 \hline
bCGC &        2457.02 (2363.15) &    2117.49 (2040.88)  \\
IP-SAT &      2468.24 (2397.26) &    2122.40 (2067.21)  \\
IPnon-SAT &       2359.12 (2284.88) &    2027.97 (1969.70)  \\
 \hline\hline
\multicolumn{3}{c}{Effective nuclear photon flux: $N^{(2)}(\omega,b)$} \\
 \hline
bCGC &        2219.46 (2138.48) &    1994.70 (1924.65)  \\
IP-SAT &       2224.47 (2165.74) &    1996.35 (1947.48)  \\
IPnon-SAT &      2124.79 (2063.21) &    1906.64 (1854.94)  \\
 \hline\hline
\multicolumn{3}{c}{Effective nuclear photon flux: $N^{(3)}(\omega,b)$} \\
 \hline
bCGC &        2220.97 (2140.55) &    1994.84 (1924.77)  \\
IP-SAT &       2225.92 (2167.87) &    1996.51 (1947.61)  \\
IPnon-SAT &       2126.18 (2065.12) &    1906.80 (1855.07)  \\
 \hline\hline
 \end{tabular}
\caption{Predictions for the ratio between total cross-sections associated with the photoproduction of $J/\Psi$ mesons in  peripheral $PbPb$ collisions at \( \sqrt{s} = 5.02\,\mathrm{TeV} \) and $OO$ collisions at  \( \sqrt{s} = 5.36\,\mathrm{TeV} \). Results for forward rapidities ($2 \leq y_{\Psi} \leq 4.5$) and different centralities,  derived considering distinct models for the effective nuclear photon flux,  dipole - proton scattering amplitude and overlap function. The predictions in parentheses are for the GLC model.}
\label{Tab:ratio_forward}
 \end{table}

\section{Summary}
\label{sec:sum}
In this paper, we investigated $J/\Psi$ photoproduction in peripheral $OO$ collisions at the LHC. We considered different approaches for the main ingredients of the calculation in order to estimate the current theoretical uncertainties. Our results indicate that, given the integrated luminosity delivered in the recent $OO$ run, an experimental analysis of this process is, in principle, feasible. Moreover, our predictions complement those obtained in Ref.~\cite{daCosta:2025frd} for $PbPb$ collisions, since both studies use the same assumptions for the effective nuclear photon flux, the dipole--proton scattering amplitude, and the overlap function. Finally, we proposed studying the ratio of the $J/\Psi$ photoproduction cross sections in peripheral $PbPb$ and $OO$ collisions as a way to improve our understanding of photon-induced interactions in peripheral collisions. { These conclusions motivate improving some of the assumptions used in our analysis, in particular the centrality definition and the treatment of nuclear geometry. Such improvements, together with the inclusion of effects associated with quarkonium propagation and dissociation in a (hot) nuclear medium, will also allow us to provide predictions for more central collisions. Both topics are currently under investigation, and we intend to present our results in forthcoming studies.}

\begin{acknowledgments}
V.P.G was partially financed by the Brazilian funding agencies CNPq, FAPERGS and INCT-FNA  (Process No. 408419/2024-5). P.E.A.C and B.D.M. were partially supported by CAPES and FAPESC (Process No. 2025TR001585). 
A.V.G acknowledges support from CNPq through the INCT-FNA grant 408419/2024-5, the 
Universal Grant 405458/2025-8 and the National Laboratory for Scientific Computing (LNCC/MCTI, Brazil), 
through the ambassador program (UFGD, subproject SADFT), for providing HPC resources of the 
SDumont supercomputer.


\end{acknowledgments}

\bibliographystyle{unsrt}

\begin{thebibliography}{99}

\bibitem{Yagi:2005yb}
K.~Yagi, T.~Hatsuda and Y.~Miake,
Camb. Monogr. Part. Phys. Nucl. Phys. Cosmol. \textbf{23}, 1-446 (2005)

\bibitem{Busza:2018rrf}
W.~Busza, K.~Rajagopal and W.~van der Schee,
Ann. Rev. Nucl. Part. Sci. \textbf{68}, 339-376 (2018)

\bibitem{upc}
C. A. Bertulani and G. Baur, { Phys. Rep.} {\bf 163}, 299 (1988); F.~Krauss, M.~Greiner and G.~Soff,
  Prog.\ Part.\ Nucl.\ Phys.\  {\bf 39}, 503 (1997);
   C.~A. Bertulani, S.~R.~Klein and J.~Nystrand, Ann. Rev. Nucl. Part. Sci. {\bf 55}, 
271 (2005); V.~P.~Goncalves and M.~V.~T.~Machado,
  J.\ Phys.\ G {\bf 32}, 295 (2006);       A.~J.~Baltz {\it et al.},
  Phys.\ Rept.\  {\bf 458}, 1 (2008);       J.~G.~Contreras and J.~D.~Tapia Takaki,
  Int.\ J.\ Mod.\ Phys.\ A {\bf 30}, 1542012 (2015); 
      K.~Akiba {\it et al.} [LHC Forward Physics Working Group],
  J.\ Phys.\ G {\bf 43}, 110201 (2016);  S.~Klein and P.~Steinberg,
Ann. Rev. Nucl. Part. Sci. \textbf{70}, 323-354 (2020).

\bibitem{klein} S. R. Klein, J. Nystrand,  Phys. Rev. C {\bf 60},
014903 (1999).




\bibitem{gluon}
  V.~P.~Goncalves and C.~A.~Bertulani,
  Phys.\ Rev.\ C {\bf 65}, 054905 (2002).



\bibitem{Frankfurt:2001db}
L.~Frankfurt, M.~Strikman and M.~Zhalov,
Phys. Lett. B \textbf{540}, 220-226 (2002).



\bibitem{upc2}
S.~R.~Klein and H.~Mantysaari,
Nature Rev. Phys. \textbf{1}, no.11, 662-674 (2019).



\bibitem{Nagle:2018nvi}
J.~L.~Nagle and W.~A.~Zajc,
Ann. Rev. Nucl. Part. Sci. \textbf{68}, 211-235 (2018)


\bibitem{Brewer:2021kiv}
J.~Brewer, A.~Mazeliauskas and W.~van der Schee,
[arXiv:2103.01939 [hep-ph]].



\bibitem{Nijs:2021clz}
G.~Nijs and W.~van der Schee,
Phys. Rev. C \textbf{106}, no.4, 044903 (2022)



\bibitem{Goncalves:2022ret}
V.~P.~Goncalves, B.~D.~Moreira and L.~Santana,
Phys. Rev. C \textbf{107}, no.5, 055205 (2023).


\bibitem{Eskola:2022vaf}
K.~J.~Eskola, C.~A.~Flett, V.~Guzey, T.~L{\"o}yt{\"a}inen and H.~Paukkunen,
Phys. Rev. C \textbf{107}, no.4, 044912 (2023)
[erratum: Phys. Rev. C \textbf{108}, no.6, 069901 (2023)]

\bibitem{Cepila:2025exl}
J.~Cepila, J.~G.~Contreras, M.~Matas and A.~Ridzikova,
Phys. Rev. C \textbf{113}, no.2, 2 (2026).





\bibitem{CMS:2025bta}
A.~Hayrapetyan \textit{et al.} [CMS],
Phys. Rev. Lett. \textbf{136}, no.16, 162301 (2026).

\bibitem{CMS:2026qef}
A.~Belyaev \textit{et al.} [CMS],
[arXiv:2602.21325 [nucl-ex]].



\bibitem{ALICE:2015mzu}
J.~Adam \textit{et al.} [ALICE],
Phys. Rev. Lett. \textbf{116}, no.22, 222301 (2016)

\bibitem{STAR:2019yox}
J.~Adam \textit{et al.} [STAR],
Phys. Rev. Lett. \textbf{123}, no.13, 132302 (2019)

\bibitem{LHCb:2021hoq}
R.~Aaij \textit{et al.} [LHCb],
Phys. Rev. C \textbf{105}, no.3, L032201 (2022)

\bibitem{ALICE:2022zso}
S.~Acharya \textit{et al.} [ALICE],
Phys. Lett. B \textbf{846}, 137467 (2023)

\bibitem{Massacrier:2024fgx}
L.~Massacrier [ALICE],
[arXiv:2407.09707 [nucl-ex]].


\bibitem{Bize:2024ros}
N.~Biz\'e [ALICE],
Phys. Proc. UPC \textbf{1}, 18 (2024)













\bibitem{Klusek-Gawenda:2015hja}
M.~K\l{}usek-Gawenda and A.~Szczurek,
Phys. Rev. C \textbf{93}, no.4, 044912 (2016)


\bibitem{Zha:2017jch}
W.~Zha, S.~R.~Klein, R.~Ma, L.~Ruan, T.~Todoroki, Z.~Tang, Z.~Xu, C.~Yang, Q.~Yang and S.~Yang,
Phys. Rev. C \textbf{97}, no.4, 044910 (2018)

\bibitem{GayDucati:2017ksh}
M.~B.~Gay Ducati and S.~Martins,
Phys. Rev. D \textbf{96}, no.5, 056014 (2017)



\bibitem{GayDucati:2018who}
M.~B.~Gay Ducati and S.~Martins,
Phys. Rev. D \textbf{97}, no.11, 116013 (2018)


\bibitem{Shi:2017qep}
W.~Shi, W.~Zha and B.~Chen,
Phys. Lett. B \textbf{777}, 399-405 (2018)


\bibitem{Zha:2018ytv}
W.~Zha, L.~Ruan, Z.~Tang, Z.~Xu and S.~Yang,
Phys. Lett. B \textbf{789}, 238-242 (2019)



\bibitem{daCosta:2025frd}
P.~E.~A.~da Costa, A.~V.~Giannini, V.~P.~Goncalves and B.~D.~Moreira,
Phys. Rev. D \textbf{112}, no.3, 034012 (2025)



\bibitem{epa} 
  V.~M.~Budnev, I.~F.~Ginzburg, G.~V.~Meledin and V.~G.~Serbo,
  Phys.\ Rept.\  {\bf 15}, 181 (1975).


\bibitem{DeJager:1974liz} 
  C.~W.~De Jager, H.~De Vries and C.~De Vries,
  Atom.\ Data Nucl.\ Data Tabl.\  {\bf 14}, 479 (1974)
  Erratum: [Atom.\ Data Nucl.\ Data Tabl.\  {\bf 16}, 580 (1975)].







  \bibitem{hdqcd} 
  F.~Gelis, E.~Iancu, J.~Jalilian-Marian and R.~Venugopalan,
    Ann.\ Rev.\ Nucl.\ Part.\ Sci.\  {\bf 60}, 463 (2010);
  H.~Weigert,  Prog.\ Part.\ Nucl.\ Phys.\  {\bf 55}, 461 (2005); J.~Jalilian-Marian and Y.~V.~Kovchegov, Prog.\ Part.\ Nucl.\ Phys.\  {\bf 56}, 104 (2006); A.~Morreale and F.~Salazar,
Universe \textbf{7}, no.8, 312 (2021).







\bibitem{glauber}
R. J. Glauber, in Lecture in Theoretical Physics, Vol. 1, edited by W. E. Brittin, L. G. Duham (Interscience, New York, 1959).



\bibitem{gribov}
V.~N.~Gribov,
Sov.\ Phys.\ JETP {\bf 29}, 483 (1969); Sov.\ Phys.\ JETP {\bf 30}, 709 (1970).
  

\bibitem{mueller}
A.~H.~Mueller,
Nucl.\ Phys.\ B {\bf 335}, 115 (1990).

\bibitem{Armesto:2002ny}
N.~Armesto,
Eur.\ Phys.\ J.\ C\ {\bf 26}, 35 (2013).


\bibitem{KMW}               H.~Kowalski, L.~Motyka and G.~Watt,  
                            Phys.\ Rev.\  D {\bf 74}, 074016 (2006). 


\bibitem{Watt_bCGC}         G.~Watt and H.~Kowalski, 
                            Phys.\ Rev.\ D {\bf 78}, 014016 (2008).






\bibitem{ipsat1}            H.~Kowalski and D.~Teaney, 
                            Phys.\ Rev.\ D {\bf 68}, 114005 (2003).


\bibitem{ipsat3}            H.~Kowalski, T.~Lappi and R.~Venugopalan,  
                            Phys.\ Rev.\ Lett.\  {\bf 100}, 022303 (2008).
  



\bibitem{ipsat4}            A.~H.~Rezaeian, M.~Siddikov, M.~Van de Klundert and R.~Venugopalan,
                            Phys.\ Rev.\ D {\bf 87}, 034002 (2013).




  \bibitem{ipsat_heikke} 
  H.~Mantysaari and P.~Zurita,
  Phys.\ Rev.\ D {\bf 98}, 036002 (2018)


\bibitem{Loizides:2017ack}
C.~Loizides, J.~Kamin and D.~d'Enterria,
Phys. Rev. C \textbf{97}, no.5, 054910 (2018)
[erratum: Phys. Rev. C \textbf{99}, no.1, 019901 (2019)]


\bibitem{Loizides:2025ule}
C.~Loizides,
Phys. Rev. C \textbf{113}, no.1, 1 (2026)

\bibitem{Mantysaari:2017slo}
H.~M{\"a}ntysaari and R.~Venugopalan,
Phys. Lett. B \textbf{781}, 664-671 (2018)

\end{thebibliography}

\end{document}